\documentclass[onecolumn]{aa}

\usepackage{amsmath}
\usepackage{eufrak}
\usepackage{txfonts}
\usepackage{graphicx}
\usepackage[colorlinks=true, linkcolor=blue, citecolor=red, urlcolor=blue]{hyperref}
\usepackage{natbib}
\bibpunct{(}{)}{;}{a}{}{,}
\pdfoutput=1

\newcommand{\vhigh}{\vphantom{\frac{1}{1}}}
\newcommand{\calC}{{\mathcal{C}}}
\newcommand{\calE}{{\mathcal{E}}}

\newcommand{\bfa}{{\boldsymbol{a}}}
\newcommand{\bfb}{{\boldsymbol{b}}}

\newcommand{\bfr}{{\boldsymbol{r}}}
\newcommand{\bfv}{{\boldsymbol{v}}}
\newcommand{\bfalpha}{{\boldsymbol{\alpha}}}
\newcommand{\bfbeta}{{\boldsymbol{\beta}}}
\newcommand{\txa}{{\text{a}}}
\newcommand{\txd}{{\text{d}}}

\newcommand{\txp}{{\text{p}}}

\renewcommand{\leq}{\leqslant}

\DeclareMathOperator{\Ei}{Ei}
\DeclareMathOperator*{\Laplace}{{\mathcal{L}}}
\DeclareMathOperator*{\Mellin}{{\mathcal{M}}}
\newenvironment{Hmatrix}{\begin{array}{r@{\,,\,}l}}{\end{array}}

\begin{document}

\title{Dynamical models with a general anisotropy profile}

\author{
M. Baes
\and
E. Van Hese
}

\authorrunning{M.~Baes \& E.~Van Hese}

\titlerunning{Dynamical models with a general anisotropy profile}

\offprints{M.~Baes,\email{maarten.baes@ugent.be}}

\institute{Sterrenkundig Observatorium, Universiteit Gent,
Krijgslaan 281 S9, B-9000 Gent, Belgium}

\date{Received / Accepted }

\abstract{} {Both numerical simulations and observational evidence
  indicate that the outer regions of galaxies and dark matter haloes
  are typically mildly to significantly radially anisotropic. The
  inner regions can be significantly non-isotropic, depending on the
  dynamical formation and evolution processes. In an attempt to break
  the lack of simple dynamical models that can reproduce this
  behaviour, we explore a technique to construct dynamical models with
  an arbitrary density and an arbitrary anisotropy profile.} {We
  outline a general construction method and propose a more practical
  approach based on a parameterized anisotropy profile. This approach
  consists of fitting the density of the model with a set of dynamical
  components, each of which have the same anisotropy profile. Using
  this approach we avoid the delicate fine-tuning difficulties other
  fitting techniques typically encounter when constructing radially
  anisotropic models.} {We present a model anisotropy profile
  that generalizes the Osipkov-Merritt profile, and that can represent
  any smooth monotonic anisotropy profile. Based on
  this model anisotropy profile, we construct a very general
  seven-parameter set of dynamical components for which the most
  important dynamical properties can be calculated analytically. We
  use the results to look for simple one-component dynamical models
  that generate simple potential-density pairs while still supporting
  a flexible anisotropy profile. We present families of Plummer and
  Hernquist models in which the anisotropy at small and large radii
  can be chosen as free parameters. We also generalize these two
  families to a three-parameter family that self-consistently
  generates the set of Veltmann potential-density pairs. These new
  analytical models are an important step forward compared to
  isotropic or Osipkov-Merritt models and can be used to generate the
  initial conditions for realistic simulations of galaxies or dark
  matter haloes.}{}


\maketitle

\section{Introduction}
\label{introduction.sec}

Numerical simulations, in particular $N$-body or hydrodynamical
simulations, have become a major tool to study the structure,
dynamics, stability and evolution of stellar systems and dark matter
haloes. While at first sight this might seem to imply that analytical
studies of spherically symmetric dynamical models have become less
interesting, actually the opposite conclusion should be made. The
construction of realistic and simple dynamical models is utterly
important, as these models act as a reference frame or starting point
from which to generate the initial conditions for numerical
simulations. In this context, it is important to stress that the
initial conditions need to be generated from the correct distribution
function $F(\bfr,\bfv)$ which completely determines the dynamical
model. \citet{2004ApJ...601...37K} have recently demonstrated that the
correct details of the velocity distribution should be taken into
account. Simple ad-hoc Jeans dynamical models, i.e.\ models in which
the kinematics are assumed to be Gaussian distributions with velocity
dispersions derived from solving the Jeans equation, are not
sufficient and can lead to erroneous conclusions. The quest for
analytical dynamical models, which are simple enough on the one hand
and realistic enough on the other hand, is hence still important.

In the past few years, many dynamical models have been proposed that
are generated by various spherically symmetric potential-density
pairs. The early-days models mainly represented systems with a
constant density core, such as the Plummer model or the isochrone
sphere \citep{1911MNRAS..71..460P, 1960AnAp...23..474H}. It now
appears that many dynamical systems such as galaxies and dark matter
haloes contain a central density cusp; also for such models a number
of representative potential-density pairs have been constructed and
distribution function have been derived \citep{1983MNRAS.202..995J,
  1990ApJ...356..359H, 1993MNRAS.265..250D, 1994AJ....107..634T,
  1994A&A...283..783H, 1996MNRAS.278..488Z, 2002A&A...393..485B,
  2004MNRAS.351...18B, 2007MNRAS.375..773B}.

Although the construction of analytical distribution functions hence
has continued to develop, most simple models still show an important
shortcoming: usually the distribution function can only be derived
under strict and unrealistic conditions on the anisotropy of the
velocity distribution. For many potential-density pairs, analytical
distribution functions have only been derived under the assumption of
isotropy. In this case, the phase-space distribution function that
supports any density profile can be found as a simple integration
using the famous Eddington formula. There are a few popular
generalizations of the simple isotropic case that lead to dynamical
models with an anisotropic velocity dispersion. Distribution functions
with a constant anisotropy profile can be generated with a formula
that is very similar to the Eddington formula
\citep{1986PhR...133..217D, 2004MNRAS.351...18B,
  2006AJ....131..782A}. Another special category are the
Osipkov-Merritt models, engineered independently by
\citet{1979PAZh....5...77O} and \citet{1985AJ.....90.1027M}. In these
essentially one-dimensional models, the distribution function depends
on the energy and angular momentum integrals only through a linear
combination of both. The anisotropy profile for these models is
peculiar in the sense that the velocity dispersion tensor is
isotropic at small radii and becomes completely radial at large
radii. The Osipkov-Merritt models were extended by
\citet{1991MNRAS.253..414C} to models in which the velocity dispersion
tensor has an arbitrary anisotropy at small radii and becomes
completely radial at large radii. The transition formulae for all
these generalizations of the Eddington formula are sufficiently simple
that analytical distribution functions of these kinds can be
constructed for many simple potential-density pairs.

Whereas these models are a step beyond isotropic dynamical models,
they still do not correspond to the anisotropy that is observed both
in numerical simulations and real galaxies. Realistic dynamical
systems typically have a tendency towards moderate to strong radial
anisotropy at large radii. Cosmological simulations generally yield
dynamical structures that are far from isotropic. Dark matter
simulations in a cosmological CDM framework typically result in haloes
in which the anisotropy gradually increases outward to levels of
$\beta\approx0.5$ at the virial radius \citep{1996MNRAS.281..716C,
  2000ApJ...539..561C, 2001ApJ...557..533F, 2004MNRAS.352..535D}. In
general, there appears to be a connection between the logarithmic
density slope $\gamma(r) = \txd\ln\rho/\txd\ln r(r)$ and $\beta(r)$,
both for pure dark matter haloes and for structures containing dark
matter and baryons \citep{2006NewA...11..333H}. This connection was
argued to be a natural consequence of the Jeans equation
\citep{2005MNRAS.363.1057D} and it was shown numerically to be an
attractor \citep{2006JCAP...05..014H}.

In hydrodynamical simulations in which gas and stars are taken into
account, galaxies also typically have a significant radial anisotropy
at large radii \citep[e.g.][]{2007MNRAS.376...39O}. Observational
dynamical evidence appears to support these trends. For example,
detailed stellar dynamical modelling of a large set of elliptical
galaxies by \citet{2000A&AS..144...53K} indicates that these galaxies
generally contain an anisotropy profile that increases from near
isotropy in the central regions towards a significant radial
anisotropy at larger radii.

The anisotropy at small radii on the other hand is more subtle and
depends largely on the dynamical processes that shape the nucleus of
the system. In particular, the influence of a supermassive black hole,
now believed to be present in nearly all hot stellar systems, can have
a lasting influence on the central anisotropy. Models that contain a
supermassive black hole that grows adiabatically by slow accretion of
material are characterized by an isotropic to slightly tangential
anisotropy profile in the central region with $\beta\sim-0.3$
\citep{1995ApJ...440..554Q}. On the other hand, models with a binary
supermassive black hole, formed by spiralling in due to dynamical
friction, typically have much more outspoken tangential anisotropy in
the central regions, with $\beta\sim-1$
\citep{1997NewA....2..533Q}. This enhanced tangential anisotropy
results from the preferential ejection of stars on radial orbits
during the hardening of the black hole binary.  Observationally, the
study of the anisotropy profile at very small radii is difficult and
hampered by both limited resolution effects and strong projection
effects. \citet{2003ApJ...583...92G} found that the most massive
galaxies are strongly biased towards tangential orbits in the
innermost regions, whereas the lower-mass galaxies have a range of
central anisotropies.

From this body of evidence it is clear that we need to extend our set
of simple isotropic or Osipkov-Merritt dynamical models to models in
which the anisotropy profile has a more general shape, typically
rising from isotropy or moderate tangential anisotropy at small radii
to significant radial anisotropy at large radii. In the general
anisotropic case, the transition formulae between density and
distribution function are rather cumbersome (although mathematically
elegant) and involve Laplace-Mellin integral transforms
\citep{1986PhR...133..217D}. The goal of the present paper is to
describe a method to construct dynamical models with an arbitrary
potential-density pair and an arbitrary anisotropy
profile. We describe the general approach and the mathematical
formulation in Section~{\ref{general.sec}}. In the next two sections
we describe a more practical approach to construct such models: in
Section~{\ref{beta.sec}} we propose a general parameterized anisotropy
profile and in Section~{\ref{components.sec}} we construct a general
seven-parameter family of dynamical components based on this
parameterized anisotropy profile for which the most important
dynamical properties can be calculated analytically. In
Section~{\ref{models.sec}} we apply the results from the previous
sections to construct a family of Plummer and Hernquist models with a
smooth anisotropy profile with arbitrary values at small and large
radii. We then extend these two families to a three-parameter family
of generally anisotropic models that support the Veltmann set of
potential-density pairs. Finally, our results are summarized in
Section~{\ref{conclusions.sec}}.

\section{Construction of general anisotropic models}
\label{general.sec}

\subsection{The augmented density formalism}

In spherical symmetry, the distribution function depends only on two
isolating integrals: the binding energy $\calE$ and the magnitude of
the angular momentum $L$ per unit mass,
\begin{gather}
  \calE 
  =
  \psi(r) - \tfrac{1}{2}\,v_r^2 - \tfrac{1}{2}\,v_T^2,
  \\
  L = r\,v_T.
\end{gather}
In these expressions $\psi(r)$ is the positive binding potential of
the system, connected to the density through Poisson's equation
\begin{equation}
  \frac{1}{r^2}\,\frac{\txd}{\txd r}
  \left(r^2\,\frac{\txd\psi}{\txd r}\right)(r)
  =
  -4\pi G\rho(r).
\end{equation}
and $v_T$ is the transverse velocity
\begin{equation}
  v_T
  =
  \sqrt{v_\theta^2+v_\varphi^2}.
\end{equation}
When we have an expression for the distribution function, all
interesting dynamical properties can be derived. In particular, the
moments of the distribution function can be calculated as
\begin{equation}
  \mu_{2n,2m}(r)
  =
  2\pi M
  \iint F(\calE,L)\,v_r^{2n}\,v_T^{2m+1}\,\txd v_r\,\txd v_T.
\label{momdf}
\end{equation}
If we want to construct dynamical models for a given density profile
in practice, it is easier to consider an alternative characterization
of a spherical dynamical model, namely the framework of the augmented
densities. An augmented density is a bivariate function
$\tilde\rho(\psi,r)$, which expresses the density as an explicit
function of both the radius $r$ and the gravitational potential
$\psi$. One can demonstrate that each augmented density function
$\tilde\rho(\psi,r)$ completely determines a distribution function
$F(\calE,L)$ and vice versa. There are several transition formulae
between these two equivalent formalisms, including an elegant
formalism based on combined Laplace-Mellin integral transforms.

One of the advantages of the augmented density formalism is that it is
straightforward to construct self-consistent dynamical models that
correspond to a given density profile. Indeed, we just have to solve
Poisson's equation, and construct any augmented density
$\tilde\rho(\psi)$ that satisfies the relation
\begin{equation}
  \rho(r) = \tilde\rho(\psi(r),r).
\label{rhoisrho}
\end{equation}
Obviously, there are always infinitely many augmented density
functions that satisfy equation~(\ref{rhoisrho}) for a given density
profile $\rho(r)$. These different dynamical models have different
moments of the distribution function, and we often are looking for
those particular dynamical models which have a certain anisotropy
profile, defined as
\begin{equation}
  \beta(r) 
  =
  1 - \frac{\sigma_T^2(r)}{2\sigma_r^2(r)}.
\end{equation}
It appears that the augmented density formalism of dynamical models is
a very useful way to achieve this goal. Indeed, another advantage of
this formalism is that the moments of the distribution function can be
derived from the augmented density by subsequent derivations and
a single integration,
\begin{equation}
    \tilde{\mu}_{2n,2m}(\psi,r)
    =
    \frac{2^{m+n}}{\sqrt{\pi}}\,
    \frac{\Gamma\left(n+\tfrac{1}{2}\right)}
    {\Gamma\left(m+n\right)}
    \int_0^\psi
    \left(\psi-\psi'\right)^{m+n-1}
    D_{r^2}^m\left[r^{2m}\tilde\rho(\psi',r)\right]
    \txd\psi',
\end{equation}
where $D_x^m$ denotes the $m\,$th differentiation with respect to $x$. In
particular, the radial and transverse velocity dispersions can be
found from the density through the relations
\begin{gather}
    \tilde\sigma_r^2(\psi,r)
    =
    \frac{\tilde\mu_{20}(\psi,r)}{\tilde\mu_{00}(\psi,r)}
    =
    \frac{1}{\tilde\rho(\psi,r)}
    \int_0^\psi \tilde\rho(\psi',r)\,\txd\psi',
    \label{gensigr:def}
    \\
    \tilde\sigma_T^2(\psi,r)
    =
    \frac{\tilde\mu_{02}(\psi,r)}{\tilde\mu_{00}(\psi,r)}
    =
    \frac{2}{\tilde\rho(\psi,r)}
    \int_0^\psi
    D_{r^2}
    \left[r^2\,\tilde\rho(\psi',r)\right]
    \txd\psi'.
\label{gensigt:def}
\end{gather}

\subsection{Construction of general anisotropic models}

Now consider an augmented density that is a separable function of
$\psi$ and $r$,
\begin{equation}
    \tilde\rho(\psi,r)
    =
    f(\psi)\,g(r).
\label{splitaugdens}
\end{equation}
For such models, the dispersion profiles read
\begin{gather}
    \tilde\sigma_r^2(\psi,r)
    =
    \frac{1}{f(\psi)}
    \int_0^\psi f(\psi')\,\txd\psi',
    \label{sigmarfg}
    \\
    \tilde\sigma_T^2(\psi,r)
    =
    \left(1+\frac{1}{2}\frac{\txd\ln g}{\txd\ln r}\right)
    \frac{2}{f(\psi)}
    \int_0^\psi f(\psi')\,\txd\psi',
\end{gather}
such that
\begin{equation}
    \beta(r)
    =
    -\frac{1}{2}\frac{\txd\ln g}{\txd\ln r}(r).
\label{betadv}
\end{equation}
For models with a separable augmented density, the radial velocity
dispersion depends only on the $\psi$-dependent part of the augmented
density, whereas the anisotropy depends only on the $r$-dependent part.

Equation~(\ref{betadv}) offers us in principle the opportunity to
construct dynamical models with an arbitrary density profile $\rho(r)$
and an arbitrary anisotropy profile. First we solve
equation~(\ref{betadv}) for $g(r)$. Next we determine the
gravitational potential $\psi(r)$ by solving Poisson's equation, we
invert this relation as $r(\psi)$ and we set
\begin{gather}
  \bar{g}(\psi) = g(r(\psi)),
  \\
  \bar\rho(\psi) = \rho(r(\psi)),
  \\
  f(\psi) = \frac{\bar\rho(\psi)}{\bar{g}(\psi)},
\end{gather}
then the augmented density
\begin{equation}
  \tilde\rho(\psi,r) 
  =
  f(\psi)\,g(r)
\end{equation}
defines the desired model.

\subsection{Practical construction}

Although we have defined a way to construct an augmented density for a
dynamical model with an arbitrary density and anisotropy profile, this
is of limited use in practice. Indeed, the quantity we really want is
the distribution function for the model. In principle, the
distribution function for each $\tilde\rho(\psi,r)$ can be recovered
with the standard Laplace-Mellin integral transform formulae. However,
the augmented density that results from an arbitrary choice of
$\rho(r)$ and $\beta(r)$ will often be too complicated to allow an
analytical integral transform. As a result, we have to compute the
distribution function by numerical means. Unfortunately, the numerical
inversion of equation~\eqref{momdf} to obtain the distribution
function from the augmented density is in general numerically
unstable. Indeed, since the density is an integration of the
distribution function over velocity space, the density is generally
much smoother than the distribution function. The inversion procedure
of determining $f$ from $\rho$ hence has the tricky job of unsmoothing
the information contained in the smoothed density. It hence comes as
no surprise that this operation is numerically unstable. For a more
precise mathematical demonstration of the unstable character of the
inversion formulae, we refer to \citet{1986PhR...133..217D}.

A more promising strategy is to determine a general parameterized
model for $\beta(r)$ that leads to a relatively simple function
$g(r)$. For each function $f(\psi)$, we can subsequently construct a
dynamical model $\tilde\rho(\psi,r)$ that has the same anisotropy
profile. Now consider a set of base functions $f_\ell(\psi)$ that are
sufficiently simple such that the augmented density
$\tilde\rho_\ell(\psi,r) = f_\ell(\psi)\,g(r)$ can be converted
analytically to a distribution function $F_\ell(\calE,L)$. It is easy
to see that any linear combination of such dynamical models will again
define a dynamical model with the same anisotropy profile since
\begin{equation}
  \sum_\ell a_\ell\,\tilde\rho_\ell(\psi,r)
  =
  \left[\sum_\ell a_\ell\,f_\ell(\psi)\right] g(r).
\end{equation}
So if we can find a suitable linear combination of the base functions
such that
\begin{equation}
  f(\psi) 
  = 
  \frac{\bar\rho(\psi)}{\bar{g}(\psi)}
  =
  \sum_\ell a_\ell\,f_\ell(\psi),
  \label{fitcondition}
\end{equation}
we immediately obtain the required distribution function
\begin{equation}
  F(\calE,L)
  =
  \sum_\ell a_\ell\,F_\ell(\calE,L).
\end{equation}
Finding the best linear combination of the components such that the
condition (\ref{fitcondition}) is satisfied can in principle be
achieved by making a formal series expansion and determine each of the
expansion coefficients for each component. In practice however, it is
more straightforward to approximate the dynamical model by a
$\chi^2$-minimization routine. This could for example be achieved by
means of quadratic programming, in which a boundary condition is set
that forces the distribution function to remain positive in the entire
phase space. Practical realizations of this technique are already in
use by various dynamical modellers \citep{1989ApJ...343..113D,
  1993MNRAS.264..712K, 1994ApJ...432..575M, 1998MNRAS.295..197G}.

The remaining steps we still have to take are hence (1) create a
general parameterized anisotropy profile $\beta(r)$ that is flexible
enough to represent an array of realistic anisotropy profiles and
that still yields a simple $g(r)$ function, (2) provide a suitable set of
base functions or components $f_\ell(\psi)$ for which, in combination
with the $g(r)$ function derived above, the corresponding distribution
function can be computed analytically. We will look into these two
steps in the next two sections.

\section{A parameterized anisotropy profile}
\label{beta.sec}

As mentioned before, very few analytical dynamical models are known
that have a realistic anisotropy profile. Most are either completely
isotropic, have a constant anisotropy or have an anisotropy profile
according to the Osipkov-Merritt type (isotropic in the centre and
completely radially anisotropic at large radii). In this paper, we aim
at the construction of spherical dynamical models with a more general
and realistic anisotropy profile. The most realistic approach would
involve an anisotropy profile that depends explicitly on the density
slope, as described in the Introduction. However, this would in
general not yield an augmented density that allows an analytical
distribution function. Instead, we will create dynamical models with
an anisotropy profile that is an explicit function of the radius
$r$. As long as the anisotropy profiles have enough freedom, this will
in practice enable us to construct dynamical models in which the
anisotropy and density slope are at least qualitatively connected. Our
specific goal is to create dynamical models in which the anisotropy
changes monotonically from an arbitrary value $\beta_0$ at the central
regions to another arbitrary value $\beta_\infty$ at large radii,
without a priori limitations on the values of $\beta_0$ or
$\beta_\infty$. To reach this goal, we need to provide a functional
form that has the desired properties at small and large radii and that
can be inverted to give a reasonable simple form for $g(r)$. Finding
such a form for $\beta(r)$ is not obvious. A straightforward candidate
that has the desired asymptotic behaviour is an exponential profile
\begin{equation}
  \beta(r) 
  =
  \beta_\infty 
  -
  (\beta_\infty-\beta_0)
  \exp\left(-\frac{r}{r_\txa}\right).
\end{equation}
Solving for equation~(\ref{betadv}) yields, apart from an arbitrary
normalization factor,
\begin{equation}
  g(r)
  =
  \left(\frac{r}{r_\txa}\right)^{-2\beta_\infty}
  \exp\left[
    -
    2\,(\beta_\infty-\beta_0)
    \Ei_1\left(\frac{r}{r_\txa}\right)
  \right].
\end{equation}
It comes as no surprise that such a model will not result in a simple
augmented density that can be inverted to an analytical distribution
function. In order to find a more suitable candidate function for
$\beta(r)$ we inspire ourselves on the Osipkov-Merritt models, where
the anisotropy profile reads
\begin{equation}
  \beta(r)
  =
  \frac{r^2}{r_\txa^2+r^2}
  =
  \frac{(r/r_\txa)^2}{1+(r/r_\txa)^2},
\end{equation}
resulting in isotropy in the central regions and complete radial
anisotropy at large radii. The resulting solution for $g(r)$ reads
\begin{equation}
  g(r)
  =
  \left(1+\frac{r^2}{r_\txa^2}\right)^{-1}.
\end{equation}
\citet{1991MNRAS.253..414C} generalized the Osipkov-Merritt models by
considering models with an anisotropy profile
\begin{equation}
  \beta(r)
  =
  \frac{\beta_0+(r/r_\txa)^2}{1+(r/r_\txa)^2}.
\end{equation}
These models also become completely radial at large radii, but at
small radii they can assume an arbitrary anisotropy $\beta_0$. If we
introduce this expression into equation (\ref{betadv}) we obtain the
relatively simple expression
\begin{equation}
  g(r)
  =
  \left(\frac{r}{r_\txa}\right)^{-2\beta_0}
  \left(1+\frac{r^2}{r_\txa^2}\right)^{\beta_0-1}.
\end{equation}
The obvious way to generalize these results such that the anisotropy
does not have to become completely radial at large radii is to
consider the profile
\begin{equation}
  \beta(r)
  =
  \frac{\beta_0+\beta_\infty(r/r_\txa)^2}{1+(r/r_\txa)^2},
\end{equation}
which corresponds to 
\begin{equation}
  g(r)
  =
  \left(\frac{r}{r_\txa}\right)^{-2\beta_0}
  \left(1+\frac{r^2}{r_\txa^2}\right)^{-(\beta_\infty-\beta_0)}.
\end{equation}
In fact, an even more general form for the anisotropy that yields a
very similar $g$ function is
\begin{equation}
  \beta(r)
  =
  \frac{\beta_0+\beta_\infty(r/r_\txa)^{2\delta}}{1+(r/r_\txa)^{2\delta}},
\label{betagen}
\end{equation}
with $\delta>0$, which corresponds to 
\begin{equation}
  g(r)
  =
  \left(\frac{r}{r_\txa}\right)^{-2\beta_0}
  \left(1+\frac{r^{2\delta}}{r_\txa^{2\delta}}\right)^{\beta_\delta},
  \label{ggen}
\end{equation}
with
\begin{equation}
  {\beta_\delta}
  =
  \frac{\beta_0-\beta_\infty}{\delta}.
\end{equation}
The couple (\ref{betagen})-(\ref{ggen}) is a suitable choice for our
goals. On the one hand this four-parameter model anisotropy contains
enough flexibility to represent any monotonically varying anisotropy
profile (with increasing or decreasing anisotropy). The parameters
have straightforward effects: $\beta_0$ and $\beta_\infty$ set the
anisotropy at small and large radii respectively, $r_\txa$ determines
the typical transition radius between the two regimes and $\delta$
sets the sharpness by which this transition takes place. On the other
hand, the resulting $g(r)$ function is sufficiently simple to lead to
analytical dynamical components for suitably chosen $f(\psi)$
functions.

\section{A library of components}
\label{components.sec}

The missing link in our approach is a set of base functions or
components with which we can construct a linear combination. We need
to look for functions $f_\ell(\psi)$ such that the augmented density
\begin{equation}
  \tilde\rho_\ell(\psi,r)
  =
  f_\ell(\psi)
  \left(\frac{r}{r_\txa}\right)^{-2\beta_0}
  \left(1+\frac{r^{2\delta}}{r_\txa^{2\delta}}\right)^{\beta_\delta},
\end{equation}
gives rise to an analytical distribution functions
$F_\ell(\calE,L)$. We will drop the subscript $\ell$ in the remainder
of this section in order not to overload the notations.

\subsection{Power law components}

The most obvious candidate components are those in which $f(\psi)$ is
a power law, 
\begin{equation}
  \tilde\rho(\psi,r)
  =
  \rho_0
  \left(\frac{\psi}{\psi_0}\right)^p
  \left(\frac{r}{r_\txa}\right)^{-2\beta_0}
  \left(1+\frac{r^{2\delta}}{r_\txa^{2\delta}}\right)^{\beta_\delta},
\label{fam1}
\end{equation}
where we limit ourselves to systems with a finite potential well
$\psi_0=\psi(0)$.  A finite total mass is obtained if
$p+2\beta_\infty>3$. In order to calculate the distribution function
corresponding to this augmented density, we apply the general
Mellin-Laplace inversion formulae. After some algebra (see
Appendix~{\ref{df1.sec}}), we find that it can be expressed as a Fox
$H$-function:
\begin{equation}
  F(\calE,L)
  =
  \frac{\rho_0}{M(2\pi\,\psi_0)^{3/2}}\,
  \frac{\Gamma(1+p)}{\delta\,\Gamma(-{\beta_\delta})}\,
  \left(\frac{\calE}{\psi_0}\right)^{p-3/2}\,
  H_{2,2}^{1,1}
  \left(
    \frac{L^2}{2r_\txa^2\calE}
    \left|
      \,\begin{Hmatrix} 
        \left(1-\frac{\beta_\infty}{\delta},\frac{1}{\delta}\right)&
        \left(p-\frac{1}{2},1\right)
        \\[2mm]
        \left(-\frac{\beta_0}{\delta},\frac{1}{\delta}\right)&
        \left(0,1\right)
      \end{Hmatrix}\, 
    \right.
  \right).
\label{dffoxH}
\end{equation}
As we show in Appendix~{\ref{df1.sec}}, a sufficient (but not
necessary) condition to yield a well-defined distribution function,
i.e.\ continuous and non-negative, is $\delta < 1$ or $\delta = 1$ and
$p-\frac{1}{2}+\beta_\delta>0$.

For practical purposes, a series expansion is more useful. After some
lengthy algebra, one obtains the fairly simple expression (see
Appendix~{\ref{df2.sec}})
\begin{equation}
  F(\calE,L)
  =
  \frac{\rho_0\,\Gamma(1+p)}{M(2\pi\,\psi_0)^{3/2}}
  \left(\frac{\calE}{\psi_0}\right)^{p-3/2}
  \times
  \begin{cases}
    \;
    \displaystyle
    \sum_{k=0}^\infty
    \binom{{\beta_\delta}}{k}\,
    \dfrac{1}{
      \Gamma\left(1-\beta_0+k\delta\right)
      \Gamma\left(p-\frac{1}{2}+\beta_0-k\delta\right)
    }\,
    \left(\dfrac{L^2}{2r_\txa^2\calE}\right)^{-\beta_0+k\delta}
    &
    \qquad
    \text{for }L^2<2r_\txa^2\calE,
    \\[1.7em]
    \;
    \displaystyle
    \sum_{k=0}^\infty
    \binom{{\beta_\delta}}{k}\,
    \dfrac{1}{
      \Gamma\left(1-\beta_\infty-k\delta\right)
      \Gamma\left(p-\frac{1}{2}+\beta_\infty+k\delta\right)
    }\,
    \left(\dfrac{L^2}{2r_\txa^2\calE}\right)^{-\beta_\infty-k\delta}
    &
    \qquad
    \text{for }L^2>2r_\txa^2\calE.
  \end{cases}
\label{dfseries}
\end{equation}
Apart from the distribution function, most other interesting dynamical
properties can be calculated analytically for this set of
components. Particularly simple are the velocity dispersions, which we
find through (\ref{sigmarfg}),
\begin{gather}
  \tilde\sigma_r^2(\psi,r)
  =
  \frac{\psi}{1+p},
  \\
  \tilde\sigma_\theta^2(\psi,r)
  =
  \tilde\sigma_\varphi^2(\psi,r)
  =
  \frac{(1-\beta_0)+(1-\beta_\infty)\,(r/r_\txa)^{2\delta}}
  {1+(r/r_\txa)^{2\delta}}\,
  \tilde\sigma_r^2(\psi,r).
\end{gather}

\subsection{Extension to cuspy components}

This set of power law components presented in the previous subsection
is very adequate to fit a broad class of models. It is however, not
fit to construct dynamical models for systems that have a central
density cusp. For such models, the isotropic distribution function
diverges in the limit $\calE\rightarrow\psi_0$ \citep[see
e.g.][]{1990ApJ...356..359H, 1993MNRAS.265..250D, 1994AJ....107..634T,
  2005A&A...432..411B}. This behaviour cannot be represented with the
current set of components. Therefore we extend the set of models
defined by the augmented density~(\ref{fam1}) to
\begin{equation}
  \tilde\rho(\psi,r)
  =
  \rho_0
  \left(\frac{\psi}{\psi_0}\right)^p\,
  \left(1-\frac{\psi^s}{\psi_0^s}\right)^q
  \left(\frac{r}{r_\txa}\right)^{-2\beta_0}
  \left(1+\frac{r^{2\delta}}{r_\txa^{2\delta}}\right)^{\beta_\delta},
\label{fam2}
\end{equation}
with two additional parameters $q\leq0$ and $s>0$. This
seven-parameter family of dynamical components contains enough freedom
in the density profile to form a good family of base functions. If we
expand this expression as a power series in $\psi$,
\begin{equation}
  \tilde\rho(\psi,r)
  =
  \rho_0
  \sum_{j=0}^\infty
  (-1)^j\,
  \binom{q}{j}
  \left(\frac{\psi}{\psi_0}\right)^{p+js}
  \left(\frac{r}{r_\txa}\right)^{-2\beta_0}
  \left(1+\frac{r^{2\delta}}{r_\txa^{2\delta}}\right)^{\beta_\delta},
\end{equation}
we see that we obtain a series of positive terms that all have the
form (\ref{fam1}). Thus, our condition $q\leq0$ is sufficient to
obtain a well-defined distribution function, which we can write down
immediately,
\begin{equation}
  F(\calE,L)
  =
  \frac{\rho_0}{M(2\pi\,\psi_0)^{3/2}}\,
  \sum_{j=0}^\infty
  (-1)^j\,
  \binom{q}{j}\,
  \frac{\Gamma(1+p+js)}{\delta\,\Gamma(-{\beta_\delta})}\,
  \left(\frac{\calE}{\psi_0}\right)^{p+js-3/2}\,
  H_{2,2}^{1,1}
  \left(
    \frac{L^2}{2r_\txa^2\calE}
    \left|
      \,\begin{Hmatrix} 
        \left(1-\frac{\beta_\infty}{\delta},\frac{1}{\delta}\right)&
        \left(p+js-\frac{1}{2},1\right)
        \\[2mm]
        \left(-\frac{\beta_0}{\delta},\frac{1}{\delta}\right)&
        \left(0,1\right)
      \end{Hmatrix}\, 
    \right.
  \right),
\label{fam2df}
\end{equation}
or explicitly
\begin{multline}
  F(\calE,L)
  =
  \frac{\rho_0}{M(2\pi\,\psi_0)^{3/2}}\,
  \sum_{j=0}^\infty
  (-1)^j\,
  \binom{q}{j}\,
  \Gamma(1+p+js)
  \left(\frac{\calE}{\psi_0}\right)^{p+js-3/2}
  \\
  \times
  \begin{cases}
    \;
    \displaystyle
    \sum_{k=0}^\infty
    \binom{{\beta_\delta}}{k}\,
    \dfrac{1}{
      \Gamma\left(1-\beta_0+k\delta\right)
      \Gamma\left(p+js-\frac{1}{2}+\beta_0-k\delta\right)
    }\,
    \left(\dfrac{L^2}{2r_\txa^2\calE}\right)^{-\beta_0+k\delta}
    &
    \qquad
    \text{for }L^2<2r_\txa^2\calE,
    \\[1.7em]
    \;
    \displaystyle
    \sum_{k=0}^\infty
    \binom{{\beta_\delta}}{k}\,
    \dfrac{1}{
      \Gamma\left(1-\beta_\infty-k\delta\right)
      \Gamma\left(p+js-\frac{1}{2}+\beta_\infty+k\delta\right)
    }\,
    \left(\dfrac{L^2}{2r_\txa^2\calE}\right)^{-\beta_\infty-k\delta}
    &
    \qquad
    \text{for }L^2>2r_\txa^2\calE.
  \end{cases}
\end{multline}
Since the summations in $j$ converge for every fixed value 
of $k$, this distribution function is indeed well-defined.
Other dynamical properties, such as the moments of the distribution
function, can be derived in the same way. For the velocity dispersions
we find
\begin{gather}
  \tilde\sigma_r^2(\psi,r)
  =
  \frac{\psi_0}{s}
  \left(\frac{\psi}{\psi_0}\right)^{-p}
  \left(1-\frac{\psi^s}{\psi_0^s}\right)^{-q}
  B_{\psi^s/\psi_0^s}\!\left(\frac{1+p}{s},1+q\right),
  \\
  \tilde\sigma_\theta^2(\psi,r)
  =
  \tilde\sigma_\varphi^2(\psi,r)  
  =
  \frac{(1-\beta_0)+(1-\beta_\infty)\,(r/r_\txa)^{2\delta}}
  {1+(r/r_\txa)^{2\delta}}\,
  \tilde\sigma_r^2(\psi,r),
\end{gather}
with $B_x(a,b)$ the incomplete Beta function.

\section{Self-consistent analytical models}
\label{models.sec}

In the previous Section we have described a practical method to
construct analytical self-consistent dynamical models that generate
any spherical density profile and with an anisotropy profile defined
by the general parameterized function (\ref{betagen}). We are applying
this technique to construct dynamical models matching the density and
anisotropy profiles that have come out of cosmological simulations of
dark matter haloes, in order to gain insight into their phase-space
structure \citep{2007prepV}. In the remainder of this current paper we
will focus on simple analytical dynamical models, by investigating
whether the global seven-parameter family of dynamical models defined
by the augmented density (\ref{fam2}) contains any simple
self-consistent models (hence without the need to make a linear
combination). Thus we will look for potential-density pairs
$[\rho(r),\psi(r)]$ that satisfy both Poisson's equation and the
relation 
\begin{equation}
  \rho(r) 
  =
  \tilde\rho(\psi(r),r),
\label{condsc}
\end{equation}
with $\tilde\rho(\psi,r)$ the seven-parameter augmented density from
equation~(\ref{fam2}). The existence of such analytical dynamical
models would be a great step forward in our quest for simple but
realistic dynamical models that can be used as a framework in which to
initiate detailed numerical simulations.

\subsection{A family of anisotropic Plummer models}

One of the most obvious candidates is the Plummer model, as this model
has a rather straightforward potential-density pair
\begin{gather}
  \psi(r)
  =
  \frac{GM}{\sqrt{b^2+r^2}},
  \\
  \rho(r)
  =
  \frac{3M}{4\pi b^3}\left(1+\frac{r^2}{b^2}\right)^{-5/2}.
\end{gather}
When we combine these functions with the expressions (\ref{fam2}) and
(\ref{condsc}), we obtain the condition
\begin{equation}
  \frac{3M}{4\pi b^3}
  \left(1+\frac{r^2}{b^2}\right)^{-5/2}
  =
  \rho_0
  \left(1+\frac{r^2}{b^2}\right)^{-p/2}
  \left[1-\left(1+\frac{r^2}{b^2}\right)^{-s/2}\right]^q
  \left(\frac{r}{r_\txa}\right)^{-2\beta_0}
  \left(1+\frac{r^{2\delta}}{r_\txa^{2\delta}}\right)^{-(\beta_\infty-\beta_0)/\delta}.
\label{plcond}
\end{equation}
A straightforward solution is obviously given by
\begin{gather}
  \rho_0 = \frac{3M}{4\pi b^3},
  \\
  p = 5,
  \\
  q = \beta_0 = \beta_\infty = 0,
\end{gather}
which yields the isotropic Plummer model, defined by 
\begin{gather}
  \tilde\rho(\psi)
  =
  \frac{3M}{4\pi b^3}\,\left(\frac{b\psi}{GM}\right)^5.
  \\
  F(\calE)
  =
  \frac{3}{7\pi^3\,(GMb)^{3/2}}
  \left(\frac{2b\calE}{GM}\right)^{7/2}.
\end{gather}
Our goal, however, is to determine the most general subspace of the
$(p,q,s,\beta_0,\beta_\infty,\delta,r_\txa)$ parameter space such that
the condition~(\ref{plcond}) is satisfied for all $r$. In particular
we aim for a subspace of the parameter space that has no restrictions
on $\beta_0$ and $\beta_\infty$, such that we obtain a family of
dynamical models with an arbitrary anisotropy at small and large
radii. It is obvious from equation~(\ref{plcond}) that, in order to
find non-trivial solutions, we have to set
\begin{gather}
  s = 2\delta = 2,
  \\
  r_\txa = b.
\end{gather}
We then obtain the expression
\begin{equation}
  \frac{3M}{4\pi b^3\rho_0}
  \left(\frac{r^2}{b^2}\right)^{\beta_0-q}
  \left(1+\frac{r^2}{b^2}\right)^{-5/2+p/2+q+\beta_\infty-\beta_0}
  =
  1.
\end{equation}
These expressions are identical for all values of $r$ if 
\begin{gather}
  \rho_0 = \frac{3M}{4\pi b^3},
  \\
  p = 5 - 2\beta_\infty,
  \\
  q = \beta_0.
\end{gather}
We now have constructed a general two-parameter family of
self-consistent Plummer models with the augmented density profile
\begin{equation}
  \tilde\rho(\psi,r)
  =
  \frac{3M}{4\pi\,b^3}
  \left(\frac{b\psi}{GM}\right)^{5-2\beta_\infty}
  \left[1-\left(\frac{b\psi}{GM}\right)^2\right]^{\beta_0}
  \left(\frac{r^2}{b^2}\right)^{-\beta_0}
  \left(1+\frac{r^2}{b^2}\right)^{-(\beta_\infty-\beta_0)}.
\end{equation}
From the condition $q\leq 0$ it follows that the central anisotropy
has to satisfy $\beta_0\leq0$.  This is in fact a special case of the
cusp slope-central anisotropy theorem of \citet{2006ApJ...642..752A}:
a model without a density cusp cannot have a radial velocity
anisotropy in the centre.  The anisotropy at large radii can take any
value $\beta_\infty\leq1$. By construction, the anisotropy profile of
this family of models reads
\begin{equation}
  \beta(r)
  =
  \frac{\beta_0b^2+\beta_\infty r^2}{b^2+r^2},
\end{equation}
which indeed tends towards $\beta_0$ at small radii and towards
$\beta_\infty$ at large radii. The radial velocity dispersion profile
can be written as
\begin{equation}
  \sigma_r^2(r)
  =
  \frac{GM}{2b}
  \left(\frac{r^2}{b^2}\right)^{-\beta_0}
  \left(1+\frac{r^2}{b^2}\right)^{5/2-(\beta_\infty-\beta_0)}
  B_{\frac{b^2}{b^2+r^2}}(3-\beta_\infty,1+\beta_0).
\end{equation}  
At large radii, the radial velocity dispersion profile shows a
$r^{-1}$ behaviour for all values of the parameters $\beta_0$ and
$\beta_\infty$,
\begin{gather}
  \sigma_r^2(r)
  \sim
  \frac{1}{6-2\beta_\infty}\,\frac{GM}{r},
\end{gather}
The asymptotic behaviour at small radii is
\begin{gather}
  \sigma_r^2(r)
  \sim
  \begin{cases}
    \;-\dfrac{GM}{b}\,\dfrac{1}{1+\beta_0}\,\dfrac{r^2}{b^2}
    &
    \qquad\text{if }\beta_0<-1,
    \\[4mm]
    \;\dfrac{GM}{2b}\,B(3-\beta_\infty,1+\beta_0)
    \left(\dfrac{r^2}{b^2}\right)^{-\beta_0}
    &
    \qquad\text{if }\beta_0>-1.
  \end{cases}
\end{gather}
Except for the models that are isotropic in the centre, the radial
velocity dispersions hence always tend to zero at small radii. The
asymptotic behaviour of the tangential velocity dispersions
$\sigma_\theta(r)=\sigma_\varphi(r)$ follows immediately. At large radii
we obtain
\begin{equation}
  \sigma_\theta^2(r)
  =
  \sigma_\varphi^2(r)
  \sim
  \frac{1-\beta_\infty}{6-2\beta_\infty}\,\frac{GM}{r},
\end{equation}
whereas at small radii
\begin{equation}
  \sigma_\theta^2(r)
  =
  \sigma_\varphi^2(r)
  \sim
  \begin{cases}
    \;-\dfrac{GM}{b}\,\dfrac{1-\beta_0}{1+\beta_0}\,\dfrac{r^2}{b^2}
    &
    \qquad\text{if }\beta_0<-1,
    \\[4mm]
    \;\dfrac{1-\beta_0}{2}\,
    \dfrac{GM}{b}\,
    B(3-\beta_\infty,1+\beta_0)
    \left(\dfrac{r^2}{b^2}\right)^{-\beta_0}
    &
    \qquad\text{if }\beta_0>-1.
  \end{cases}
\end{equation}
\begin{figure}
\centering
\includegraphics[width=\textwidth]{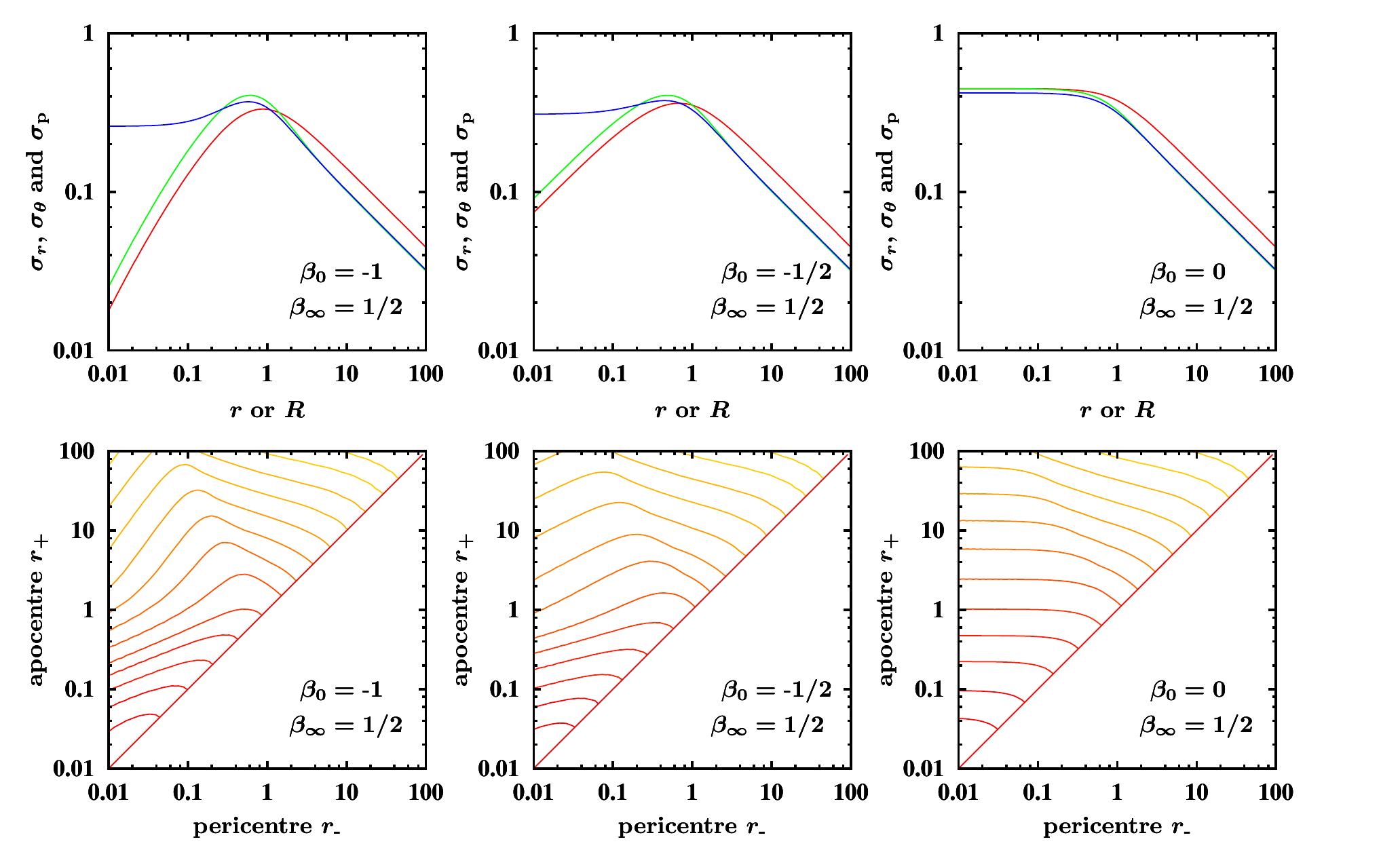}
\caption{ \emph{Top.}  Radial (red), tangential (green) and projected
  (blue) velocity dispersion profiles for three Plummer models with
  anisotropy values $\beta_0$ and $\beta_\infty$ displayed in the
  figures. \emph{Bottom.}  The distribution function of these models,
  represented by isoprobability contours in turning point space. High
  values are indicated by red contours, low values by yellow
  contours. In all plots, we have used normalized units with $G = M =
  b = 1$.}
\label{plummer.pdf}
\end{figure}
In the top panels of Fig.~{\ref{plummer.pdf}} we plot the radial and
tangential velocity dispersions for a set of anisotropic Plummer
models. The three models in this figure all have the same 
anisotropy $\beta_\infty = \tfrac{1}{2}$ at large radii, but a
different central anisotropy. Apart from the radial and transversal
velocity dispersion, we also plot the projected velocity dispersion
$\sigma_\txp(R)$ on the plane of the sky, defined through
\begin{equation}
  \rho_\txp(R)\,\sigma_\txp^2(R)
  =
  2\int_R^\infty
  \left[1-\frac{R^2}{r^2}\,\beta(r)\right]
  \frac{\rho(r)\,\sigma_r^2(r)\,r\,\txd r}{\sqrt{r^2-R^2}}.
\end{equation}
Here, $\rho_\txp(R)$ is the projected mass density on the plane of the
sky,
\begin{equation}
  \rho_\txp(R)
  =
  2\int_R^\infty
  \frac{\rho(r)\,r\,\txd r}{\sqrt{r^2-R^2}}
  =
  \frac{M}{\pi}\,
  \frac{b^2}{(b^2+R^2)^2}.
\end{equation}
The projected velocity dispersion always reaches a finite value in the
centre and has an $R^{-1}$ dependence at large radii,
\begin{equation}
  \sigma_\txp^2(R)
  \sim
  \frac{3\pi}{128}\,
  \frac{6-5\beta_\infty}{3-\beta_\infty}\,\frac{GM}{R}.
\end{equation}
The distribution function for our set of Plummer models can be written
as
\begin{equation}
  F(\calE,L)
  =
  \frac{3}{2(2\pi)^{5/2}}\,\frac{1}{(GMb)^{3/2}}\,
  \sum_{j=0}^\infty
  (-1)^j\,
  \binom{\beta_0}{j}\,
  \frac{\Gamma(6+2j-2\beta_\infty)}{\Gamma(\beta_\infty-\beta_0)}
  \left(\frac{b\calE}{GM}\right)^{7/2+2j-2\beta_\infty}
  {\mathbb{H}}
  \left(-\beta_0,\beta_\infty,\frac{9}{2}+2j-2\beta_\infty,1;
    \frac{L^2}{2b^2\calE}\right),
\label{genpldf}
\end{equation}
where the function $\mathbb{H}(a,b,c,d;x)$ is a function defined as
\citep{1986PhR...133..217D}
\begin{equation}
  \mathbb{H}(a,b,c,d;x)
  =
  G_{2,2}^{1,1}
  \left(
    x
    \left|
      \,\begin{matrix} 1-b,c \\ a,1-d \end{matrix}\, 
    \right.
  \right)
  \equiv
  H_{2,2}^{1,1}
  \left(
    x
    \left|
      \,\begin{matrix} (1-b,1),(c,1) \\ 
      (a,1),(1-d,1) \end{matrix}\, 
    \right.
  \right),
\end{equation}
with $G^{m,n}_{p,q}(z)$ the Meijer $G$-function. This
$\mathbb{H}$-function can conveniently be expressed as
\begin{equation}
  \mathbb{H}(a,b,c,d;x)
  =
  \begin{cases}
    \;
    \dfrac{\Gamma(a+b)}{\Gamma(c-a)\,\Gamma(a+d)}\,
    x^a\,{}_2F_1(a+b,1+a-c;a+d;x)
    &
    \qquad\text{if }x<1,
    \\[4mm]
    \;
    \dfrac{\Gamma(a+b)}{\Gamma(d-b)\,\Gamma(b+c)}\,
    \left(\dfrac{1}{x}\right)^b
    {}_2F_1\left(a+b,1+b-d;b+c;\dfrac{1}{x}\right)
    &
    \qquad\text{if }x>1.
  \end{cases}
\end{equation}
In the bottom panels of Fig.~{\ref{plummer.pdf}} we plot the
distribution function as a contour plot in the turning point
space for the same models as the upper panels. The change in
anisotropy from tangentially anisotropic at small radii to radially
anisotropic at large radii can easily be seen in the slope of these
contours.

An interesting subset of models in our two-parameter family of Plummer
models is the one-parameter family with $\beta_0=0$. These models are
isotropic in the inner regions and become anisotropic at large
radii. Their augmented density is given by
\begin{equation}
  \tilde\rho(\psi,r)
  =
  \frac{3M}{4\pi\,b^3}
  \left(\frac{b\psi}{GM}\right)^{5-2\beta_\infty}
  \left(1+\frac{r^2}{b^2}\right)^{-\beta_\infty}.
\end{equation}
and in the expression for the distribution function (\ref{genpldf}) only
the term corresponding to $j=0$ remains
\begin{equation}
  F(\calE,L)
  =
  \frac{3}{2(2\pi)^{5/2}}\,
  \frac{1}{(GMb)^{3/2}}
  \frac{\Gamma(6-2\beta_\infty)}{\Gamma(\beta_\infty)}
  \left(\frac{b\calE}{GM}\right)^{7/2-2\beta_\infty}
  {\mathbb{H}}
  \left(0,\beta_\infty,\frac{9}{2}-2\beta_\infty,1;
    \frac{L^2}{2b^2\calE}\right).
\end{equation}
This subfamily of our current set of Plummer models was already
presented by \citet{1987MNRAS.224...13D}. Most of the kinematical
properties, including the projected properties such as dispersions and
higher-order moments of the line profiles, can be calculated
completely analytically.

\subsection{A family of anisotropic Hernquist models}

Another extremely popular and simple potential-density pair is the
Hernquist model, defined by
\begin{gather}
  \psi(r)
  =
  \frac{GM}{b+r},
  \label{herpot}
  \\
  \rho(r)
  =
  \frac{M}{2\pi}\,
  \frac{b}{r\left(b+r\right)^3}.
  \label{herdens}
\end{gather}
Contrary to the Plummer model, this model has a central $r^{-1}$
density cusp and a more realistic $r^{-4}$ behaviour at large
radii. We can do the same experiment for the Hernquist model as we did
for the Plummer model. If we combine the potential-density pair
(\ref{herpot})-(\ref{herdens}) with expression~(\ref{fam2})
and~(\ref{condsc}), we obtain
\begin{equation}
  \frac{M}{2\pi b^3}
  \left(\frac{r}{b}\right)^{-1}
  \left(1+\frac{r}{b}\right)^{-3}
  =
  \rho_0
  \left(1+\frac{r}{b}\right)^{-p}
  \left[1-\left(1+\frac{r}{b}\right)^{-s}\right]^q
  \left(\frac{r}{r_\txa}\right)^{-2\beta_0}
  \left(1+\frac{r^{2\delta}}{r_\txa^{2\delta}}\right)^{-(\beta_\infty-\beta_0)/\delta}.
\label{hercond}
\end{equation}
Similarly as for the Plummer model, it is clear that we will only be
able to find a general non-trivial solution if we set
\begin{gather}
  s = 2\delta = 1,
  \\
  r_\txa = b.
\end{gather}
This yields the equation
\begin{equation}  
  \frac{M}{2\pi b^3\rho_0}
  \left(\frac{r}{b}\right)^{-1-q+2\beta_0}
  \left(1+\frac{r}{b}\right)^{-3+p+q+2\beta_\infty-2\beta_0}
  =
  1,
\end{equation}
from which we find
\begin{gather}
  \rho_0 = \frac{M}{2\pi b^3},
  \\
  p = 4 - 2\beta_\infty,
  \\
  q = 2\beta_0 - 1.
\end{gather}
\begin{figure}
\centering
\includegraphics[width=\textwidth]{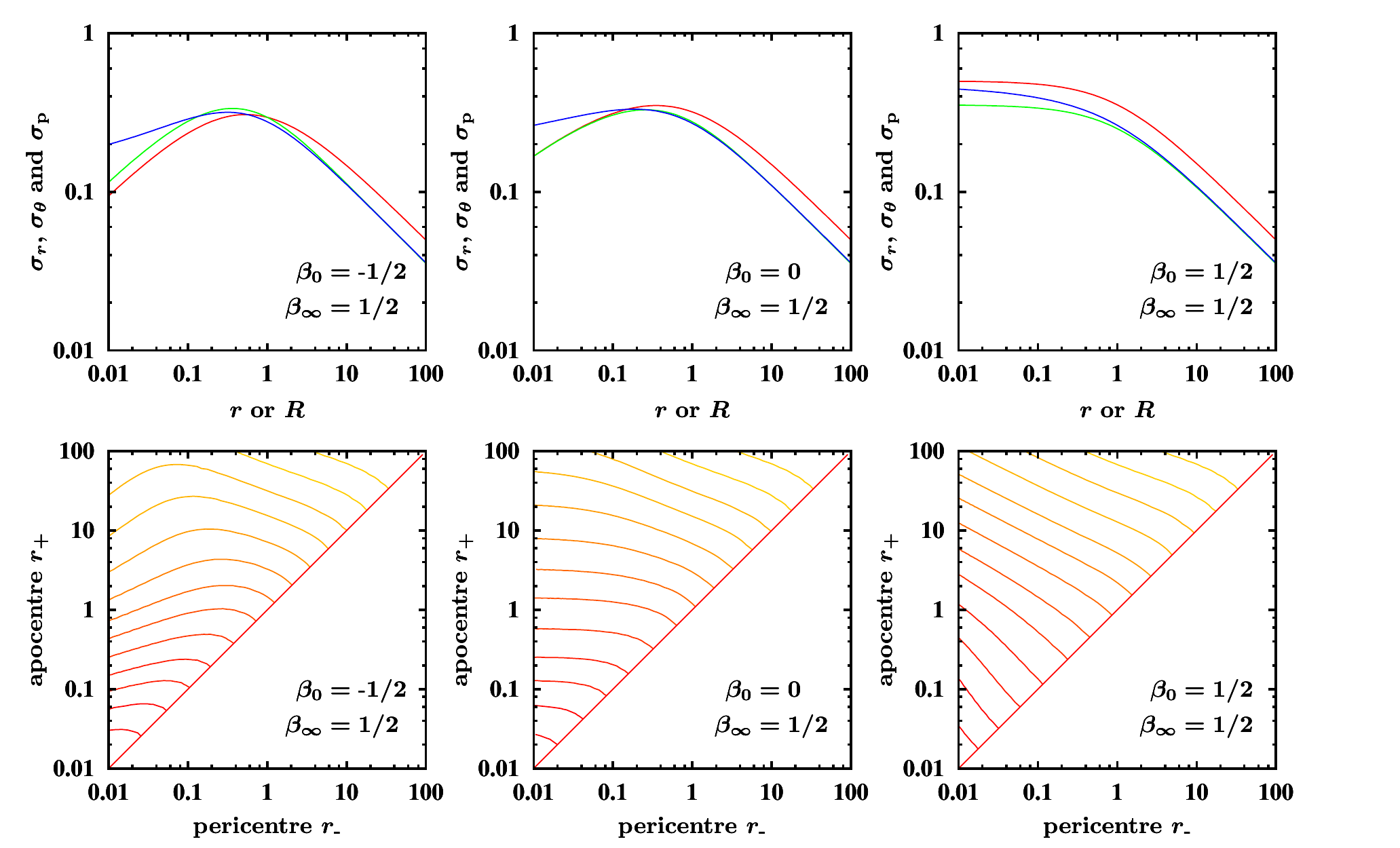}
\caption{ \emph{Top.}  Radial (red), tangential (green) and projected
  (blue) velocity dispersion profiles for three Hernquist models with
  anisotropy values $\beta_0$ and $\beta_\infty$ displayed in the
  figures.  \emph{Bottom.}  The distribution function of these models,
  represented by isoprobability contours in turning point space. High
  values are indicated by red contours, low values by yellow
  contours. In all plots, we have used normalized units with $G = M =
  b = 1$.}
\label{hernquist.pdf}
\end{figure}
We have now defined a two-parameter family of self-consistent
Hernquist models with augmented density
\begin{equation}
  \tilde\rho(\psi,r)
  =
  \frac{M}{2\pi b^3}\,
  \left(\frac{b\psi}{GM}\right)^{4-2\beta_\infty}
  \left(1-\frac{b\psi}{GM}\right)^{2\beta_0-1}
  \left(\frac{r}{b}\right)^{-2\beta_0}
  \left(1+\frac{r}{b}\right)^{-2(\beta_\infty-\beta_0)}.
\end{equation}
The parameter $\beta_\infty$ can assume all values, whereas the
central anisotropy $\beta_0$ is limited to $\beta_0\leq\tfrac{1}{2}$,
in agreement with the cusp slope-central anisotropy theorem of
\citet{2006ApJ...642..752A}. By construction, the anisotropy profile
of this family of Hernquist models reads
\begin{equation}
  \beta(r)
  =
  \frac{\beta_0b+\beta_\infty r}{b+r},
\end{equation}
which has the desired behaviour at small and large radii. The radial
dispersion profile reads
\begin{equation}
  \sigma_r^2(r)
  =
  \frac{GM}{b}
  \left(\frac{r}{b}\right)^{1-2\beta_0}
  \left(1+\frac{r}{b}\right)^{3-2(\beta_\infty-\beta_0)}
  B_{\frac{b}{b+r}}(5-2\beta_\infty,2\beta_0).
\end{equation}
At large radii, the radial velocity dispersion profile falls as $r^{-1}$,
\begin{gather}
  \sigma_r^2(r)
  \sim
  \frac{1}{5-2\beta_\infty}\,\frac{GM}{r},
\end{gather}
whereas the asymptotic behaviour at small radii is
\begin{gather}
  \sigma_r^2(r)
  \sim
  \begin{cases}
    \;-\dfrac{GM}{b}\,\dfrac{1}{2\beta_0}\,\dfrac{r}{b}
    &
    \qquad\text{if }\beta_0<0,
    \\[4mm]
    \;
    \dfrac{GM}{b}\,
    B(5-2\beta_\infty,2\beta_0)
    \left(\dfrac{r}{b}\right)^{1-2\beta_0}
    &
    \qquad\text{if }\beta_0>0.
  \end{cases}
\end{gather}
The radial velocity dispersion hence always disappears in the centre,
except for the models with the largest allowed central anisotropy
($\beta_0=\tfrac{1}{2}$) where it reaches a finite value. For the
asymptotic behaviour of the tangential velocity dispersions
$\sigma_\theta(r)=\sigma_\varphi(r)$ at large radii we obtain
\begin{equation}
  \sigma_\theta^2(r)
  =
  \sigma_\varphi^2(r)
  \sim
  \frac{1-\beta_\infty}{5-2\beta_\infty}\,\frac{GM}{r},
\end{equation}
whereas at small radii
\begin{equation}
  \sigma_\theta^2(r)
  =
  \sigma_\varphi^2(r)
  \sim
  \begin{cases}
    \;-\dfrac{GM}{b}\,\dfrac{1-\beta_0}{2\beta_0}\,\dfrac{r}{b}
    &
    \qquad\text{if }\beta_0<0,
    \\[4mm]
    \;\dfrac{GM}{b}\,
    (1-\beta_0)\,B(5-2\beta_\infty,2\beta_0)
    \left(\dfrac{r}{b}\right)^{1-2\beta_0}
    &
    \qquad\text{if }\beta_0>0.
  \end{cases}
\end{equation}
The projected velocity dispersion on the plane of the sky has a
$R^{-1}$ dependence at large radii,
\begin{equation}
  \sigma_\txp^2(R)
  \sim
  \frac{8}{15\pi}\,
  \frac{5-4\beta_\infty}{5-2\beta_\infty}\,\frac{GM}{R}.
\end{equation}
The distribution function can most conveniently be written as a series
of hypergeometric functions
\begin{multline}
  F(\calE,L)
  =
  \frac{1}{(2\pi)^{5/2}}\,
  \frac{1}{(GMb)^{3/2}}\,
  \left(\frac{b\calE}{GM}\right)^{5/2-2\beta_\infty}
  \\
  \times
  \sum_{k=0}^\infty 
  \binom{{\beta_\delta}}{k}\,
  \frac{\Gamma(5-2\beta_\infty)}
  {\Gamma\left(1-\beta_0+\frac{k}{2}\right)
    \Gamma\left(\tfrac{7}{2}-2\beta_\infty+\beta_0-\frac{k}{2}\right)}
  \left(\frac{L^2}{2b^2\calE}\right)^{-\beta_0+k/2}\,
  {}_2F_1\left(
    5-2\beta_\infty,
    1-2\beta_0;
    \frac{7}{2}-\beta_\infty+\frac{k}{2};
    \frac{b\calE}{GM}
  \right)
\end{multline}
if $L^2<2b^2\calE$, and as
\begin{multline}
  F(\calE,L)
  =
  \frac{1}{(2\pi)^{5/2}}\,
  \frac{1}{(GMb)^{3/2}}\,
  \left(\frac{b\calE}{GM}\right)^{5/2-2\beta_\infty}
  \\
  \times
  \sum_{k=0}^\infty 
  \binom{{\beta_\delta}}{k}\,
  \frac{\Gamma(5-2\beta_\infty)}
  {\Gamma\left(1-\beta_\infty-\frac{k}{2}\right)
    \Gamma\left(\tfrac{7}{2}-\beta_\infty+\frac{k}{2}\right)}
  \left(\frac{L^2}{2b^2\calE}\right)^{-\beta_\infty-k/2}\,
  {}_2F_1\left(
    5-2\beta_\infty,
    1-2\beta_0;
    \frac{7}{2}-2\beta_\infty+\beta_0-\frac{k}{2};
    \frac{b\calE}{GM}
  \right)
\end{multline}
if $L^2>2b^2\calE$. These sums only contain a finite number of terms if
${\beta_\delta}$ is a positive integer number, i.e.\ when
$(\beta_0-\beta_\infty)$ is a positive integer or half-integer
number. This particular subset of models, in which the outer regions
are always more tangentially anisotropic than the central regions, has
already been discussed by \citet{2002A&A...393..485B}.

In a similar manner as for the Plummer model, we plot in
Fig.~\ref{hernquist.pdf} the velocity dispersions and the distribution
function for three Hernquist models with the same anisotropy
$\beta_\infty$ but different anisotropy $\beta_0$.

\subsection{Generalization to a three-parameter family of anisotropic
  Veltmann models}
  
It is well-known that the Plummer and the Hernquist potential-density
pairs can be generalized to a one-parameter family of models
characterized by
\begin{gather}
  \psi(r)
  =
  \frac{GM}{(b^\lambda+r^\lambda)^{1/\lambda}},
  \label{veltpot}
  \\
  \rho(r)
  =
  \frac{(1+\lambda)\,M}{4\pi}\,
  \frac{b^\lambda}{r^{2-\lambda}\,(b^\lambda+r^\lambda)^{2+1/\lambda}}.
  \label{veltrho}
\end{gather}
This potential-density pair was first described by
\citet{1979AZh....56..976V} and is a special subset (the
$\alpha$-models) of the general set of potential-density pairs
considered by \citet{1996MNRAS.278..488Z}. This potential-density pair
recently regained much interest because it supports dynamical models
that are hypervirial, i.e.\ in which the virial relation is not only
satisfied on a global but also on a local level
\citep{2005MNRAS.360..492E, 2006PhRvE..73d6112I,
  2006PThPS.162...62S}. The parameter $\lambda$, lying in the range
$0<\lambda\leq2$, determines the slope of the central density cusp. We
easily recognize the Plummer model with $\lambda=2$ as the only
core-density member of the family and the Hernquist model as the model
with $\lambda=1$.
  
We can now repeat the same exercise as for the Plummer and Hernquist
models. After a little bit of algebra, we find that the parameters
\begin{gather}
  \rho_0 = \frac{(1+\lambda)\,M}{4\pi b^3},
  \label{veltpar1}
  \\
  p = 3 + \lambda - 2\beta_\infty,
  \\
  q = 1 + \frac{2\left(\beta_0-1\right)}{\lambda},
  \label{veltpar3}
  \\
  s = 2\delta = \lambda,
  \\
  r_\txa = b,
  \label{veltpar5}
\end{gather}
are the general solution for the condition of self-consistency.
Notice that the initial condition $q \leq 0$ implies $\beta_0 \leq 1 -
\lambda/2$, which is in correspondence with the cusp slope-central
anisotropy theorem \citep{2006ApJ...642..752A}. In other words, for
this family the condition $q \leq 0$ is also necessary to yield
physical models. In this manner we have constructed a three-parameter
family of dynamical models defined by the augmented density
\begin{equation}
  \tilde\rho(\psi,r)
  =
  \frac{(1+\lambda)\,M}{4\pi b^3}
  \left(\frac{b\psi}{GM}\right)^{3+\lambda-2\beta_\infty}
  \left[1-\left(\frac{b\psi}{GM}\right)^\lambda\right]^{1+2(\beta_0-1)/\lambda}
  \left(\frac{r^\lambda}{b^\lambda}\right)^{-2\beta_0/\lambda}
  \left(1+\frac{r^\lambda}{b^\lambda}\right)^{-2(\beta_\infty-\beta_0)/\lambda}.
\end{equation}
This augmented density self-consistently supports the one-parameter
potential-density pair (\ref{veltpot})-(\ref{veltrho}) and has by
construction an anisotropy profile
\begin{equation}
  \beta(r)
  =
  \frac{\beta_0b^\lambda+\beta_\infty r^\lambda}{b^\lambda+r^\lambda},
\end{equation}
which varies smoothly from $\beta_0$ in the centre towards
$\beta_\infty$ at large radii. The radial velocity dispersion can be
written as
\begin{equation}
  \sigma_r^2(r)
  =
  \frac{GM}{\lambda b}
  \left(\frac{r^\lambda}{b^\lambda}\right)^{-1+2(1-\beta_0)/\lambda}
  \left(1+\frac{r^\lambda}{b^\lambda}\right)^{2+(1-2(\beta_\infty-\beta_0))/\lambda}
  B_{\frac{b^\lambda}{b^\lambda+r^\lambda}}
  \left(1+\frac{2\,(2-\beta_\infty)}{\lambda},
    2-\frac{2\,(1-\beta_0)}{\lambda}\right).
\end{equation}  
For general values of $\lambda$, the distribution function cannot be
simplified and should be taken as in formula~(\ref{fam2df}) with the
values (\ref{veltpar1})-(\ref{veltpar5}). For rational values of
$\lambda$ however, distribution function simplifies to a sum of
generalized hypergeometric functions. One such case is the model with
$\lambda=1/2$, for which we obtain the potential-density pair
\begin{gather}
  \psi(r)
  =
  \frac{GM}{\left(\sqrt{b}+\sqrt{r}\right)^2},
  \\
  \rho(r)
  =
  \frac{3M}{8\pi}\,
  \frac{\sqrt{b}}{r^{3/2}\,\left(\sqrt{b}+\sqrt{r}\right)^4}.
\end{gather}
This model has a central density cups with a slope $\rho(r) \propto
r^{-3/2}$, which has been obtained by numerical simulations for dark
matter haloes in a CDM cosmological model \citep{1998ApJ...499L...5M}.

\section{Discussion and conclusions}
\label{conclusions.sec}

In this paper, we have described a method to construct self-consistent
spherical dynamical models that have an arbitrary density and an arbitrary
anisotropy profile. While the general formulation is not so
useful since the inversion formulae are rather complicated and
numerically unstable, we propose an approach with a parameterized
anisotropy profile. We put forward a very general four-parameter
anisotropy profile in which the four parameters control the anisotropy
at the centre and at large radii, the typical transition radius
between the two regimes and the sharpness in which this transition
takes place. This parameterized function can therefore represent a
large number of monotonically varying anisotropy profiles.

Based on these anisotropy profiles, we present a seven-parameter set
of dynamical components that all have the same anisotropy. For each of
these components the augmented density $\tilde\rho(\psi,r)$ is a
sufficiently simple function so that the corresponding distribution
function can be written analytically (as a double series). For many
reasonable choices of the parameters, the distribution function can be
written as a sum of generalized hypergeometric series. The lower-order
moments can be expressed by simple analytical functions -- for
example, the velocity dispersions of all components can be expressed
as incomplete Beta functions.

Once the anisotropy profile has been fixed, the construction of a
dynamical model with a given potential-density pair consists of a
determination of the best linear combination of the various
components. This can be achieved with a quadratic programming
approach, a technique already in use by various dynamical modellers
\citep{1989ApJ...343..113D, 1993MNRAS.264..712K, 1994ApJ...432..575M,
  1998MNRAS.295..197G}. This dynamical modelling technique is rather
advanced: dynamical models can be constructed by fitting a variety of
data, such as any moments of the distribution function, the projected
moments along the line of sight, the full line profiles or even
complete spectra. One could argue what the novelty of our approach is
compared to the general powerful techniques already presented by these
authors. The use of the quadratic programming technique is very
general and does not require all components to have the same
anisotropy profile. Any possible combination of components can in
principle be combined. The main strategy usually consists in choosing
the components in such a way that the distribution functions are
simple enough; one can for example choose only Fricke components which
are double power laws both in augmented density and distribution
function. As long as the data base of possible models is large enough,
such components allow to fit basically any dynamical model since a
linear combination of two components with a different density profile
and different constant anisotropy results in a model with a
non-constant anisotropy. However, practice has learned that it is
generally difficult to construct models with a strong radial
anisotropy at large radii. The reason for this problem is that one
needs to populate the model with radial orbits that reach large
radii. Since such orbits also contribute to the density at small
radii, it requires a delicate fine-tuning of the different components
to both satisfy the density and anisotropy constraints at small radii
while still retaining the radial anisotropy at large radii. In our
present case, this problem does not surface since we only have to fit
the function $f(\psi)$, or equivalently the density profile. Any
combination of the base components will automatically have the correct
anisotropy profile. This was actually one of the motivations for the
construction of this set of dynamical components.

Apart from presenting an approach to construct detailed dynamical
models with a given density and anisotropy profile, our study has also
lead to a set of simple one-component dynamical models. The models we
have generated are self-consistent realizations of the Plummer and
Hernquist models. We have also generalized these models to a complete
set of models that self-consistently generate the family of Veltmann
(or generalized Plummer) potential-density pairs. Contrary to most
analytical dynamical models that have appeared in the literature, the
anisotropy profile of these new dynamical models can be chosen
arbitrarily in the sense that we can choose the anisotropy at both
small and large radii. Both numerical simulations and observational
evidence clearly indicates that the outer regions of galaxies and dark
matter haloes are typically mildly to significantly radially
anisotropic and that the inner regions can be significantly
non-isotropic. Realistic simulations should use this information and
not impose unrealistic constraints on the anisotropy. We therefore
believe that these new analytical models are an important step forward
compared to the isotropic or Osipkov-Merritt type dynamical models
that are often used. We encourage numerical astrophysicists to use
these new distribution functions to generate the initial conditions
for their simulations of galaxies or dark matter haloes. Numerical
implementation of the most important dynamical properties of these
models can be obtained from the authors.

We are currently employing the modelling technique described in this
paper to construct dynamical models for dark matter haloes
\citep{2007prepV}. We are using constraints on the density and the
anisotropy profile of these haloes from detailed $N$-body
simulations. This will be useful to investigate the phase space
structure of dynamical models and will also provide the community with
an advanced toy model for dark haloes from which more simulations can
be initiated. This nicely illustrates that the rise of numerical
simulations does not erase the need for analytical work, but rather
that there is a symbiosis between both approaches.

\begin{appendix}
\section{Derivation of the distribution function}
\label{df1.sec}

We now present the detailed calculation of the distribution function
corresponding with the augmented density~\eqref{fam1}. We have to
consider several cases, depending on the parameter $\beta_\delta$.

\subsection{Case~1: $\beta_\delta$ is a natural number}

First, we consider the special case where $\beta_\delta$ is zero or a
natural number. Then the augmented density is simply a finite sum of
positive Fricke components
\citep{1952AN....280..193F,1973A&A....24..229H}
\begin{equation}
  \tilde\rho(\psi,r)
  =
  \rho_0
  \left(\frac{\psi}{\psi_0}\right)^p
  \sum_{k=0}^{\beta_\delta}\binom{\beta_\delta}{k}
  \left(\frac{r}{r_\txa}\right)^{-2\beta_0+2k\delta},
\label{rhosum1}
\end{equation}
which leads immediately to equation~\eqref{dfseries}. 

\subsection{Case~2: $\beta_\delta<0$}

When $\beta_\delta<0$ we can apply the Laplace-Mellin formalism: the
connection between the distribution function and the augmented density
is then \citep{1986PhR...133..217D}
\begin{equation}
  \Laplace_{\calE\rightarrow\xi}\,\Mellin_{L\rightarrow\eta}
  \,\{F(\calE,L)\}
  = \frac{2^{\eta/2}}{M(2\pi)^{3/2}}
  \frac{\xi^{(3-\eta)/2}}{\Gamma\left(1 - \frac{\eta}{2}\right)}
  \,\Laplace_{\psi\rightarrow\xi}\,\Mellin_{r\rightarrow\eta}
  \,\{\tilde\rho(\psi,r)\}.
\end{equation}
Since the augmented density is a separable function of $\psi$ and $r$,
the transforms can be calculated separately:
\begin{equation}
  \Laplace_{\psi\rightarrow\xi}\,
  \left\{ \psi^p\right\}
  = 
  \int_0^\infty e^{-\xi\psi}\psi^p\,\txd\psi
  = 
  \frac{\Gamma(1+p)}{\xi^{1+p}},
\end{equation}
and
\begin{equation}
  \Mellin_{r\rightarrow\eta}\,
  \left\{
    \left(\frac{r}{r_\txa}\right)^{-2\beta_0}
    \left(1+\frac{r^{2\delta}}{r_\txa^{2\delta}}\right)^
    {\beta_\delta}
  \right\}
  = 
  \int_0^\infty r^{\eta-1} 
  \left(\frac{r}{r_\txa}\right)^{-2\beta_0}
  \left(1+\frac{r^{2\delta}}{r_\txa^{2\delta}}\right)^
  {-(\beta_\infty-\beta_0)/\delta}\txd r
  = 
  \frac{r_a^\eta}{2\delta} 
  B\left(\frac{\eta-2\beta_0}{2\delta},\frac{2\beta_\infty-\eta}
    {2\delta}\right),
\end{equation}
where $\eta$ lies in the convergence strip $2\beta_0 < \eta <
2\beta_\infty$. Thus
\begin{equation}
  \Laplace_{\calE\rightarrow\xi}\,\Mellin_{L\rightarrow\eta}
  \,\{F(\calE,L)\}
  = 
  \frac{\rho_0 2^{\eta/2}}{M\psi_0^p(2\pi)^{3/2}} \frac{r_a^\eta}{2\delta} 
  \frac{\Gamma(1+p)}{\Gamma\left(1 - \frac{\eta}{2}\right)}
  B\left(\frac{\eta-2\beta_0}{2\delta},\frac{2\beta_\infty-\eta}{2\delta}\right)
  \xi^{(1-\eta)/2-p}.
\end{equation}
The inversion of the Laplace transform is straightforward,
\begin{equation}
  {\Laplace_{\xi\rightarrow \calE}}^{-1}
  \,\left\{\xi^{(1-\eta)/2-p}\right\}
  = 
  \frac{1}{2\pi i}\int_{\xi_0-i\infty}^{\xi_0+i\infty}
  e^{\calE\xi}\xi^{(1-\eta)/2-p}\,\txd\xi
  =
  \frac{\calE^{p-(3-\eta)/2}}{\Gamma\left(p - \frac{1-\eta}{2}\right)},
\end{equation}
leaving us with the inversion of the Mellin transform
\begin{align}
  F(\calE,L) 
  &= 
  \frac{\rho_0}{M(2\pi\psi_0)^{3/2}}
  \frac{\Gamma(1+p)}
  {\delta\,\Gamma(-\beta_\delta)} 
  \left(\frac{\calE}{\psi_0}\right)^{p-3/2}
  {\Mellin_{\eta\rightarrow L}}^{-1}
  \,\left\{\frac{1}{2}
    \frac{\Gamma\left(\frac{\eta-2\beta_0}{2\delta}\right)\,
      \Gamma\left(\frac{2\beta_\infty-\eta}{2\delta}\right)}
    {\Gamma\left(1 - \frac{\eta}{2}\right)\,
      \Gamma\left(p-\frac{1-\eta}{2}\right)}
    \,\left(2r_\txa^2\calE\right)^{\eta/2}\right\},
  \nonumber\\
  &= 
  \frac{\rho_0}{M(2\pi\psi_0)^{3/2}}
  \frac{\Gamma(1+p)}
  {\delta\,\Gamma(-\beta_\delta)}
  \left(\frac{\calE}{\psi_0}\right)^{p-3/2}
  \frac{1}{2\pi i} \int_C
  \frac{\Gamma\left(\frac{s-\beta_0}{\delta}\right)\,
    \Gamma\left(\frac{\beta_\infty-s}{\delta}\right)}
  {\Gamma\left(1 - s \vhigh\right)\,
    \Gamma\left(p-\frac{1}{2}+s\right)}
  \,\left(\frac{L^2}{2r_\txa^2\calE}\right)^{-s}\txd s.
\end{align}
Taking into account the definition of the Fox $H$-function, 
\begin{equation}
  H_{p,q}^{m,n}
  \left(
    z\left|
      \,\begin{matrix} 
        \left(\bfa,\bfalpha\right) 
        \\ 
        \left(\bfb,\bfbeta\right) 
      \end{matrix}\, 
    \right.
  \right)
  =
  \frac{1}{2\pi i}
  \int_\calC
  \frac{
    \prod_{j=1}^m \Gamma(b_j+\beta_js)\,
    \prod_{j=1}^n \Gamma(1-a_j-\alpha_js)
  }{
    \prod_{j=n+1}^p \Gamma(a_j+\alpha_js)\,
    \prod_{j=m+1}^q \Gamma(1-b_j-\beta_js)
  }\,
  z^{-s}\,
  \txd s,
\end{equation}
the distribution function can be written as
\begin{equation}
  F(\calE,L)
  =
  \frac{\rho_0}{M(2\pi\,\psi_0)^{3/2}}\,
  \frac{\Gamma(1+p)}{\delta\,\Gamma(-{\beta_\delta})}\,
  \left(\frac{\calE}{\psi_0}\right)^{p-3/2}\,
  H_{2,2}^{1,1}
  \left(
    \frac{L^2}{2r_\txa^2\calE}
    \left|
      \,\begin{Hmatrix} 
        \left(1-\frac{\beta_\infty}{\delta},\frac{1}{\delta}\right)&
        \left(p-\frac{1}{2},1\right)
        \\[2mm]
        \left(-\frac{\beta_0}{\delta},\frac{1}{\delta}\right)&
        \left(0,1\right)
      \end{Hmatrix}\, 
    \right.
  \right),
\label{dffoxHap}
\end{equation}
or equivalently
\begin{equation}
  F(\calE,L)
  =
  \frac{\rho_0}{M(2\pi\,\psi_0)^{3/2}}\,
  \frac{\Gamma(1+p)}{\Gamma(-{\beta_\delta})}\,
  \left(\frac{\calE}{\psi_0}\right)^{p-3/2}\,
  H_{2,2}^{1,1}
  \left(
    \left(\frac{L^2}{2r_\txa^2\calE}\right)^\delta
    \left|
      \,\begin{Hmatrix}
        \left(1-\frac{\beta_\infty}{\delta},1\right)&
        \left(p-\frac{1}{2},\delta\right)
        \\[2mm]
        \left(-\frac{\beta_0}{\delta},1\right)&
        \left(0,\delta\right)
      \end{Hmatrix}\, 
    \right.
  \right).
\end{equation}
The integration can be performed along three possible paths $C$, which are
equivalent: if the integral converges for more than one of these three 
paths, then the result is the same. If the integral converges for 
only one path, then that is the only one to be considered. The valid
contours are
\begin{itemize}
\item{$C$ is a path $C_1$ from $-i\infty$ to $i\infty$, such that the
    poles of $\Gamma\left(\frac{s-\beta_0}{\delta}\right)$ lie on one
    side and the poles of
    $\Gamma\left(\frac{\beta_\infty-s}{\delta}\right)$ lie on the
    other side. The convergence is absolute if $\delta < 1$ or if
    $\delta = 1$ and $p-\frac{1}{2}+\beta_\delta>0$.  This condition
    is found by investigating the more familiar criterion for Meijer
    $G$-functions (see Appendix~\ref{df2.sec}). In addition, if $\delta = 1$ and
    $p+\frac{1}{2}+\beta_\delta>0$ the integral is semi-convergent for
    $L^2 \neq 2r_\txa^2\calE$ 
    \citep[][Appendix]{1986PhR...133..217D}. }
\item{$C$ is a loop $C_2$, starting and ending at $-\infty$, that
    encircles the poles of
    $\Gamma\left(\frac{s-\beta_0}{\delta}\right)$ once in the positive
    direction and none of the poles of
    $\Gamma\left(\frac{\beta_\infty-s}{\delta}\right)$. The integral
    then converges if $L^2 < 2r_\txa^2\calE$.}
\item{$C$ is a loop $C_3$, starting and ending at $+\infty$, that
    encircles the poles of
    $\Gamma\left(\frac{\beta_\infty-s}{\delta}\right)$ once in the
    negative direction and none of the poles of
    $\Gamma\left(\frac{s-\beta_0}{\delta}\right)$. The integral then
    converges if $L^2 > 2r_\txa^2\calE$.}
\end{itemize}
It can easily be seen that the convergence criterion for path $C_1$
ensures a well-defined distribution function, i.e.\ continuous and
non-negative. Indeed, in this case the contour is a line parallel to
the imaginary axis from $s_0 -i\infty$ to $s_0 +i\infty$ with
$\beta_0<s_0<\beta_\infty$. On this path the real part of the Gamma
functions is positive (where the condition $p+2\beta_\infty>3$
ensures that $p-\frac{1}{2}+s>0$ by choosing $s_0$ sufficiently close
to $\beta_\infty$). Hence, the distribution function is non-negative
everywhere.

\subsection{Case 3: $\beta_\delta>0$ and not a natural number}

In the case of $\beta_\delta > 0$ and not a natural number, the Mellin
transform does not exist.  However, we can solve this problem by
rewriting the augmented density in a similar way as
equation~\eqref{rhosum1}: defining $n$ as the smallest natural number such
that $n > \beta_\delta$, we obtain
\begin{equation}
  \tilde\rho(\psi,r)
  =
  \rho_0
  \left(\frac{\psi}{\psi_0}\right)^p
  \sum_{k=0}^n\binom{n}{k}\,
  \left(\frac{r}{r_\txa}\right)^{-2\beta_0+2k\delta}
  \left(1+\frac{r^{2\delta}}{r_\txa^{2\delta}}\right)^{\beta_\delta-n}.
\end{equation}
For the individual terms the Laplace-Mellin formalism does apply,
and the distribution function becomes
\begin{equation}
  F(\calE,L) 
  = 
  \frac{\rho_0}{M(2\pi\psi_0)^{3/2}}
  \frac{\Gamma(1+p)}
  {\delta\,\Gamma(n-\beta_\delta)}
  \left(\frac{\calE}{\psi_0}\right)^{p-3/2}
  \sum_{k=0}^n\binom{n}{k}\,
  \frac{1}{2\pi i}\int_{C_{(k)}}
  \frac{\Gamma\left(\frac{s-\beta_0}{\delta}+k\right)\,
    \Gamma\left(\frac{\beta_\infty-s}{\delta}+n-k\right)}
  {\Gamma\left(1 - s \vhigh\right)\,
    \Gamma\left(p-\frac{1}{2}+s\right)}
  \,\left(\frac{L^2}{2r_\txa^2\calE}\right)^{-s}\txd s.
\end{equation}
Now, it is easily observed that we can choose the integration paths
$C_{(k)}$ to be identical, as the contours $C_1$, $C_2$ or $C_3$
defined above.  Therefore the summation can be performed inside the
integral.  Using the properties
\begin{equation}
  \Gamma(x+k) 
  = 
  (x)_k\,\Gamma(x)
  \qquad\mbox{and}\qquad
  \Gamma(x-k) = (-1)^k\frac{\Gamma(x)}{(x)_k},
\end{equation}
with $(x)_k = x(x+1)\cdots(x+k-1)$ the Pochhammer symbol, we find
\begin{equation}
  \sum_{k=0}^n\binom{n}{k}\,
  \Gamma\left(\frac{s-\beta_0}{\delta}+k\right)\,
  \Gamma\left(\frac{\beta_\infty-s}{\delta}+n-k\right) =
  \Gamma\left(\frac{s-\beta_0}{\delta}\right)\,
  \Gamma\left(\frac{\beta_\infty-s}{\delta}+n\right)\,
  {}_2F_1\left(-n,\frac{s-\beta_0}{\delta},
    \frac{s-\beta_\infty}{\delta}+1-n;1\right).
\end{equation}
Finally, with the identity
\begin{equation}
  {}_2F_1(-n,b,c;1) = \frac{(c-b)_n}{(c)_n},
\end{equation}
the equation for the distribution function also reduces to 
equation~\eqref{dffoxHap}.

If the convergence criterion for path $C_1$ is valid, then every
contour $C_{(k)}$ can be taken as a line $s_k -i\infty$ to $s_k
+i\infty$ with $\beta_0-k\delta<s_k<\beta_\infty+(n-k)\delta$.  Again,
for each $k$ the real part of the integrand is positive, so that the
distribution function is well-defined.
\end{appendix}
\begin{appendix}
\section{A practical series expansion for the distribution function}
\label{df2.sec}

We now seek a more practical form for the distribution function. To
this aim we first consider the special case where $\delta$ is a
rational number, denoting $\delta=\frac{m}{n}$. Then we can write
\begin{align}
  F(\calE,L) 
  &= 
  \frac{\rho_0}{M(2\pi\psi_0)^{3/2}}
  \frac{n\,\Gamma(1+p)}{\Gamma(-\beta_\delta)}  
  \left(\frac{\calE}{\psi_0}\right)^{p-3/2}
  \frac{1}{2\pi i}\int_C
  \frac{\Gamma\left(-\frac{n}{m}\beta_0+ns\right)\,
    \Gamma\left(\frac{n}{m}\beta_\infty-ns\right)}
  {\Gamma\left(1 - ms\vhigh\right)\,
    \Gamma\left(p-\frac{1}{2}+ms\right)}
  \left(\frac{L^2}{2r_\txa^2\calE}\right)^{-ms}\txd s.
\end{align}
Now, using the multiplication formula for the Gamma function
\begin{equation}
  \Gamma(kx) 
  = 
  (2\pi)^{(1-k)/2}k^{kx-1/2}\prod_{l=0}^{k-1}
  \Gamma\left(x+\frac{l}{k}\right), 
\label{multigam}
\end{equation}
we can write the integral in the form of a Meijer $G$-function:
\begin{equation}
  F(\calE,L) 
  = 
  \frac{\rho_0}{M(2\pi\psi_0)^{3/2}}
  \frac{1}{(2\pi)^{n-m}}
  \frac{n^{n(\beta_\infty-\beta_0)/m}}{m^{p-1/2}}
  \frac{\Gamma(1+p)}{\Gamma(-\beta_\delta)}
  \left(\frac{\calE}{\psi_0}\right)^{p-3/2}
  G_{m+n,m+n}^{n,n}
  \left(
    \left(\frac{L^2}{2r_\txa^2\calE}\right)^m
    \left|
      \,\begin{matrix} 
        \bfa 
        \\[1mm] 
        \bfb 
      \end{matrix}\, 
    \right.
  \right),
  \label{dfmeijerap}
\end{equation}
with 
\begin{align}
  a_i &= -\frac{i-1}{n} + \frac{m-\beta_\infty}{m}& &
  \mbox{for } i = 1,\ldots, n,\nonumber\\
  a_{n+i} &= \hphantom{+}\frac{i-1}{m} + \frac{p-1/2}{m}&&
  \mbox{for } i = 1,\ldots, m,\nonumber\\
  b_i &= \hphantom{+}\frac{i-1}{n} - \frac{\beta_0}{m}&&
  \mbox{for } i = 1,\ldots, n,\nonumber\\
  b_{n+i} &= -\frac{i-1}{m} + \frac{m-1}{m}&&
  \mbox{for } i = 1,\ldots, m. 
  \label{coefs}
\end{align}
Since for a general Meijer $G$-function 
\begin{equation}
  G_{p,q}^{m,n}
  \left(z
    \left|
      \,\begin{matrix} 
        \bfa 
        \\[1mm] 
        \bfb 
      \end{matrix}\, 
    \right.
  \right),
\end{equation}
with $p=q$ and $z,\bfa,\bfb$ real, the convergence criterion
for path $C_1$ states that \citep{1993Mathai}
\begin{equation}
  m+n - \frac{1}{2}(p+q) > 0
\end{equation}
or
\begin{align}
  m+n - \frac{1}{2}(p+q) &= 0,\\
  \sum_{i=0}^{q}b_i - \sum_{i=0}^{p}a_i &< -1,
\end{align}
we indeed obtain the conditions $\delta<1$ or $\delta = 1$ and
$p-\frac{1}{2}+\beta_\delta>0$, as
used in Appendix~\ref{df1.sec}.

This Meijer $G$-function can be calculated as a sum of generalized
hypergeometric functions \citep[][eqs.~9.303 and
9.304]{1965tisp.book.....G}.  For $z<1$ the following equation is
valid:
\begin{align}
  G_{m+n,m+n}^{n,n}\left(
    z^m
    \left|
      \,\begin{matrix} 
        \bfa 
        \\[1mm] 
        \bfb 
      \end{matrix}\, 
    \right.
  \right)
  =&
  \sum_{i=1}^{n}
  \frac{\displaystyle
    \prod_{l=1}^{n}{\vphantom{\prod}}^{\!\prime} \Gamma\,\bigl(b_l-b_i\bigr) 
    \prod_{l=1}^{n} \Gamma\,\bigl(1+b_i-a_l\bigr)}
  {\displaystyle\prod_{l=n+1}^{m+n} \Gamma\,\bigl(1+b_i-b_l\bigr)
    \prod_{l=n+1}^{m+n} \Gamma\,\bigl(a_l-b_i\bigr)}
  \,z^{m b_i}\nonumber\\
  \times&\; {}_{m+n}F_{m+n-1}\,\bigl(
  1+b_i-a_1,\ldots,1+b_i-a_{m+n}\,;\,\bigr.\nonumber\\
  & \hphantom{\; {}_{m+n}F_{m+n-1}\,\bigl(}
  \bigl.1+b_i-b_1,\ldots,*,\ldots,1+b_i-b_{m+n}\,;\,
  (-1)^{m+n}z^m\,\bigr)\,,
  \label{dfhyper1ap}
\end{align}
where the prime by the product symbol denotes the omission
of the product when $i=l$, and the asterisk in the hypergeometric
function indicates the omission on the $i\,$th parameter.
Analogously, the equation for $z>1$ reads
\begin{align}
  G_{m+n,m+n}^{n,n}\left(
    z^m
    \left|
      \,\begin{matrix} 
        \bfa 
        \\[1mm] 
        \bfb 
      \end{matrix}\, 
    \right.
  \right)
  =&
  \sum_{i=1}^{n}
  \frac{\displaystyle 
    \prod_{l=1}^{n}{\vphantom{\prod}}^{\!\prime} \Gamma\,\bigl(a_i-a_l\bigr) 
    \prod_{l=1}^{n} \Gamma\,\bigl(1+b_l-a_i\bigr)}
  {\displaystyle \prod_{l=n+1}^{m+n} \Gamma\,\bigl(1+a_l-a_i\bigr)
    \prod_{l=n+1}^{m+n} \Gamma\,\bigl(a_i-b_l\bigr)}
  \,z^{m (a_i-1)}\nonumber\\
  \times&\; {}_{m+n}F_{m+n-1}\,\bigl(
  1+b_1-a_i,\ldots,1+b_{m+n}-a_i\,;\,\bigr.\nonumber\\
  & \hphantom{\; {}_{m+n}F_{m+n-1}\,\bigl(}
  \bigl.1+a_1-a_i,\ldots,*,\ldots,1+a_{m+n}-a_i\,;\,
  (-1)^{m+n}z^{-m}\,\bigr)\,.
  \label{dfhyper2ap}
\end{align}
With the coefficients from equation~\eqref{coefs}, we obtain for 
$L^2<2r_\txa^2\calE$
\begin{align}
  G_{m+n,m+n}^{n,n}
  \left(
    \left(\frac{L^2}{2r_\txa^2\calE}\right)^m
    \left|
      \,\begin{matrix} 
        \bfa 
        \\[1mm] 
        \bfb 
      \end{matrix}\, 
    \right.
  \right)
  =&
  \sum_{i=0}^{n-1}
  \frac{\displaystyle
    \prod_{l=0}^{n-1}{\vphantom{\prod}}^{\!\prime}\Gamma\left(\frac{l-i}{n}\right)
    \,\prod_{l=0}^{n-1} \Gamma\left(\frac{\beta_\infty-\beta_0}{m}
      +\frac{l+i}{n}\right)}
  {\displaystyle\prod_{l=0}^{m-1}\Gamma\left(\frac{1-\beta_0+l}{m}+\frac{i}{n}\right)
    \,\prod_{l=0}^{m-1}\Gamma\left(\frac{p+\beta_0-1/2+l}{m}-\frac{i}{n}\right)}
  \nonumber\\
  \times &
  \sum_{j=0}^{\infty}
  \frac{\displaystyle
    \prod_{l=0}^{n-1}\left(\frac{\beta_\infty-\beta_0}{m}+\frac{l+i}{n}\right)_j
    \,\prod_{l=0}^{m-1}\left(1 - \frac{p+\beta_0-1/2+l}{m} + \frac{i}{n}\right)_j}
  {\displaystyle \prod_{l=0}^{n-1}
    \left(1-\frac{l-i}{n}\right)_j
    \,\prod_{l=0}^{m-1}\left(\frac{1-\beta_0+l}{m}+\frac{i}{n}\right)_j}
  (-1)^{(m+n)j}
  \left(\frac{L^2}{2r_\txa^2\calE}\right)^{-\beta_0+i\delta+mj}.
\end{align}
Now, with the aid of the identities
\begin{equation}
  \Gamma(x)\,\Gamma(1-x)
  =
  \frac{\pi}{\sin(\pi x)},
\end{equation}
and
\begin{equation}
  \prod_{k=1}^{n-1}
  \sin\left(\frac{k\pi}{n}\right) 
  = 
  \frac{n}{2^{n-1}},
\end{equation}
we can simplify this expression to
\begin{align}
  G_{m+n,m+n}^{n,n}
  \left(
    \left(\frac{L^2}{2r_\txa^2\calE}\right)^m
    \left|
      \,\begin{matrix} 
        \bfa 
        \\[1mm] 
        \bfb 
      \end{matrix}\, 
    \right.
  \right)
  =&
  \sum_{i=0}^{n-1}
  \sum_{j=0}^{\infty}
  \frac{\displaystyle
    \prod_{l=0}^{n-1} \Gamma\left(\frac{\beta_\infty-\beta_0}{m}
      +\frac{l+i}{n}+j\right)}
  {\displaystyle
    \prod_{l=0}^{n-1}\Gamma\left(\frac{1+l+i}{n}+j\right)
    \,\prod_{l=0}^{m-1}\Gamma\left(\frac{1-\beta_0+l}{m}+\frac{i}{n}+j\right)
    \,\prod_{l=0}^{m-1}\Gamma\left(\frac{p+\beta_0-1/2+l}{m}-\frac{i}{n}-j\right)}
  \nonumber\\
  \times &
  \frac{(2\pi)^{n-1}}{n}
  (-1)^{i+nj}
  \left(\frac{L^2}{2r_\txa^2\calE}\right)^{-\beta_0+i\delta+mj},
\end{align}
so that, using again equation~\eqref{multigam}, the distribution function
reduces to
\begin{align}
  F(\calE,L) 
  =& 
  \frac{\rho_0}{M(2\pi\psi_0)^{3/2}}
  \frac{\Gamma(1+p)}
  {\Gamma(-\beta_\delta)}   
  \left(\frac{\calE}{\psi_0}\right)^{p-3/2}\nonumber\\
  \times &
  \sum_{i=0}^{n-1}
  \sum_{j=0}^{\infty}
  \frac{
    \Gamma\left(\frac{\beta_\infty-\beta_0}{\delta}
      +i+jn\right)}
  {\Gamma\left(1+i+jn\right)\,
    \Gamma\left(1-\beta_0+i\delta+jm\right)\,
    \Gamma\left(p+\beta_0-1/2-i\delta-jm\right)}
  (-1)^{i+nj}\left(\frac{L^2}{2r_\txa^2\calE}\right)^{-\beta_0+i\delta+mj}.
\end{align}
Finally, the double summation can be grouped into a single index $k=i+nj$,
and we obtain for $L^2<2r_\txa^2\calE$
\begin{equation}
  F(\calE,L) 
  = 
  \frac{\rho_0}{M(2\pi\psi_0)^{3/2}}
  \left(\frac{\calE}{\psi_0}\right)^{p-3/2}
  \sum_{k=0}^{\infty}
  \binom{\beta_\delta}{k}
  \frac{\Gamma(1+p)}
  {\Gamma(1-\beta_0+k\delta)\,\Gamma(p+\beta_0-1/2-k\delta)}
  \left(\frac{L^2}{2r_\txa^2\calE}\right)^{-\beta_0+k\delta}.
\end{equation}
Similarly, for $L^2>2r_\txa^2\calE$, we find
\begin{equation}
  F(\calE,L) 
  = 
  \frac{\rho_0}{M(2\pi\psi_0)^{3/2}}
  \left(\frac{\calE}{\psi_0}\right)^{p-3/2}
  \sum_{k=0}^{\infty}
  \binom{\beta_\delta}{k}
  \frac{\Gamma(1+p)}
  {\Gamma(1-\beta_\infty-k\delta)\,\Gamma(p+\beta_\infty-1/2+k\delta)}
  \left(\frac{L^2}{2r_\txa^2\calE}\right)^{-\beta_\infty-k\delta}.
\end{equation}
Although these expressions have been derived for rational values of
$\delta$, they can be generalized to any real value, since these
functions are continuous in $\delta$. Hence we indeed obtain
equation~\eqref{dfseries}.
\end{appendix}

\end{document}